\documentclass[conference]{IEEEtran}

\usepackage{amsmath, amssymb, bm, cite, epsfig, psfrag}
\usepackage{epstopdf}
\usepackage{graphicx}
\usepackage{algorithm,algpseudocode}
\usepackage{array}
\usepackage[ampersand]{easylist}

\usepackage{bbm}
\usepackage{multirow}
\usepackage[usenames,dvipsnames]{xcolor}
\renewcommand{\arraystretch}{1.5}
\usepackage{subfigure}
\usepackage{etoolbox}
\usepackage{pbox}
\usepackage{colortbl}
\usepackage{fancyhdr}
\usepackage[margin=0.55in,bmargin=0.6in]{geometry}
\usepackage[hyphens]{url}
\usepackage[hidelinks,bookmarks=false]{hyperref}
\usepackage{cite}
\usepackage[vertfit,hyphenbreaks]{breakurl}
\usepackage[width=0.5\textwidth,font=small]{caption}
\usepackage{capt-of}
\newtoggle{conference}
\togglefalse{conference} % Use for arxiv version -- also change documentclass
\graphicspath{{figures/}}

\def\beq{\begin{equation}}
\def\eeq{\end{equation}}
\def\beqa{\begin{eqnarray}}
\def\eeqa{\end{eqnarray}}
\def\beqan{\begin{eqnarray*}}
\def\eeqan{\end{eqnarray*}}

\def\PL{\mathrm{PL}}
\def\dB{\mathrm{dB}}

\def\SF{\mathrm{SF}}

\def\tm1{t\! - \! 1}
\def\tp1{t\! + \! 1}

\def\PL{\textrm{PL}}
\def\dB{\textrm{dB}}
\def\FSPL{\textrm{FSPL}}

\def\CIF{\textrm{CIF}}

\def\ABG{\textrm{ABG}}

\def\1m{\textrm{1 m}}

\def\PL{\mathrm{PL}}
\def\dB{\mathrm{dB}}

\usepackage{lipsum}
\usepackage{fancyhdr}
\usepackage{datetime}

\usepackage{tikz}
\usetikzlibrary{calc}

\begin{document}

\newcommand\blfootnote[1]{%
  \begingroup
  \renewcommand\thefootnote{}\footnote{#1}%
  \addtocounter{footnote}{-1}%
  \endgroup
}

\pagestyle{empty}

%%---------------------- Document Header --------------------------%
%\begin{tikzpicture} [remember picture, overlay]
%\node at ($(current page.north) + (0,-0.25in)$) {M. K. Samimi, T. S. Rappaport, ``28 GHz Millimeter-Wave Ultrawideband Small-Scale Fading Models in Wireless Channels,''};
%\node at ($(current page.north) + (0,-0.4in)$) {\textit{submitted to the 2016 IEEE Vehicular Technology Conference (VTC2016-Spring)}, 15-18 May, 2016.};
%\end{tikzpicture}

\title{Indoor 5G 3GPP-like Channel Models for Office and Shopping Mall Environments}

\author{
	\IEEEauthorblockN{Katsuyuki Haneda\textsuperscript{a}, Lei Tian\textsuperscript{b}, Henrik Asplund\textsuperscript{d}, Jian Li\textsuperscript{e}, Yi Wang\textsuperscript{e}, David Steer\textsuperscript{e}, Clara Li\textsuperscript{f}, \\
		Tommaso Balercia\textsuperscript{f}, Sunguk Lee\textsuperscript{g}, YoungSuk Kim\textsuperscript{g}, Amitava Ghosh\textsuperscript{h}, Timothy Thomas\textsuperscript{h} Takehiro Nakamura\textsuperscript{i},\\
		Yuichi Kakishima\textsuperscript{i}, Tetsuro Imai\textsuperscript{i}, Haralabos Papadopoulas\textsuperscript{i}, Theodore S. Rappaport\textsuperscript{j}, George R. MacCartney Jr.\textsuperscript{j},\\
		 Mathew K. Samimi\textsuperscript{j}, Shu Sun\textsuperscript{j}, Ozge Koymen\textsuperscript{k}, Sooyoung Hur\textsuperscript{l}, Jeongho Park\textsuperscript{l}, Jianzhong Zhang\textsuperscript{l}, Evangelos Mellios\textsuperscript{m}, \\
		 Andreas F. Molisch\textsuperscript{n}, Saeed S. Ghassamzadeh\textsuperscript{o}, and Arun Ghosh\textsuperscript{o} \\
	\text{\textsuperscript{a}Aalto University}, \textsuperscript{b}BUPT, \textsuperscript{c}CMCC, \textsuperscript{d}Ericsson, \textsuperscript{e}Huawei,  \textsuperscript{f}Intel, \textsuperscript{g}KT Corporation, \textsuperscript{h}Nokia, \textsuperscript{i}NTT DOCOMO,} 
	\text{\textsuperscript{j}NYU WIRELESS, \textsuperscript{k}Qualcomm, \textsuperscript{l}Samsung, \textsuperscript{m}University of Bristol, \textsuperscript{n}University of Southern California, \textsuperscript{o}AT\&T}
}

\maketitle
\begin{tikzpicture} [remember picture, overlay]
\node at ($(current page.north) + (0,-0.25in)$) {K. Haneda \emph{et al.}, ``Indoor 5G 3GPP-like Channel Models for Office and Shopping Mall Environments,'' to be published in };
\node at ($(current page.north) + (0,-0.4in)$) {\textit{2016 IEEE International Conference on Communications Workshops (ICCW)}, May, 2016.};
\end{tikzpicture}
\begin{abstract}
Future mobile communications systems are likely to be very different to those of today with new service innovations driven by increasing data traffic demand, increasing processing power of smart devices and new innovative applications. To meet these service demands the telecommunications industry is converging on a common set of 5G requirements which includes network speeds as high as 10 Gbps, cell edge rate greater than 100 Mbps, and latency of less than 1 msec. To reach these 5G requirements the industry is looking at new spectrum bands in the range up to 100 GHz where there is spectrum availability for wide bandwidth channels. For the development of new 5G systems to operate in bands up to 100 GHz there is a need for accurate radio propagation models which are not addressed by existing channel models developed for bands below 6 GHz. This paper presents a preliminary overview of the 5G channel models for bands up to 100 GHz in indoor offices and shopping malls, derived from extensive measurements across a multitude of bands. These studies have found some extensibility of the existing 3GPP models (e.g. 3GPP TR36.873) to the higher frequency bands up to 100 GHz. The measurements indicate that the smaller wavelengths introduce an increased sensitivity of the propagation models to the scale of the environment and show some frequency dependence of the path loss as well as increased occurrence of blockage. Further, the penetration loss is highly dependent on the material and tends to increase with frequency. The small-scale characteristics of the channel such as delay spread and angular spread and the multipath richness is somewhat similar over the frequency range, which is encouraging for extending the existing 3GPP models to the wider frequency range. Further work will be carried out to complete these models, but this paper presents the first steps for an initial basis for the model development.  
\end{abstract}

\section{Introduction}
Next generation 5G cellular systems will encompass frequencies from around 500 MHz all the way to around 100 GHz. For the development of new 5G systems to operate in bands above 6 GHz, there is a need for accurate radio propagation models for these bands which are not fully modeled by existing channel models below 6 GHz, as previous generations were designed and evaluated for operation at frequencies only as high as 6 GHz. One important example is the recently developed 3D-Indoor Hotspot (InH) channel model~\cite{3GPP36873}. This paper is a summary of key results provided in a much more detailed white paper by the authors found at the link in~\cite{5GSIG}, in addition to a 3GPP-style outdoor contribution in~\cite{5GSIG_VTC16}. The 3GPP 3D channel model provides additional flexibility for the elevation dimension, thereby allowing modeling two dimensional antenna systems, such as those that are expected in next generation system deployments. It is important for future system design to develop a new channel model that will be validated for operation at higher frequencies (e.g., up to 100 GHz) and that will allow accurate performance evaluation of possible future technical specifications in indoor environments. Furthermore, the new models should be consistent with the models below 6 GHz. In some cases, the requirements may call for deviations from the modeling parameters or methodology of the existing models, but these deviations should be kept to a bare minimum and only introduced when necessary for supporting the 5G simulation use cases.

There are many existing and ongoing campaign efforts worldwide targeting 5G channel measurements and modeling. They include METIS2020~\cite{METIS2015}, COST2100/COST~\cite{COST2100}, IC1004~\cite{COSTic1004}, ETSI mmWave~\cite{ETSI2015}, NIST 5G mmWave Channel Model Alliance~\cite{NIST}, MiWEBA~\cite{MiWEBA2014}, mmMagic~\cite{mmMagic}, and NYU WIRELESS~\cite{Rap13a,Rap15a,Rap15b,Mac15a}. METIS2020, for instance, has focused on 5G technologies and has contributed extensive studies in terms of channel modelling. Their target requirements include a wide range of frequency bands (up to 86 GHz), very large bandwidths (hundreds of MHz), fully three dimensional and accurate polarization modelling, spherical wave modelling, and high spatial resolution. The METIS channel models consist of a map-based model, stochastic model, and a hybrid model which can meet requirement of flexibility and scalability.

The COST2100 channel model is a geometry-based stochastic channel model (GSCM) that can reproduce the stochastic properties of multiple-input/multiple output (MIMO) channels over time, frequency, and space. On the other hand, the 5G mmWave Channel Model Alliance is newly established and will establish guidelines for measurement calibration and methodology, modeling methodology, as well as parameterization in various environments and a database for channel measurement campaigns. NYU WIRELESS has conducted and published extensive urban propagation measurements at 28, 38, 60, and 73 GHz for both outdoor and indoor channels, and has created large-scale and small-scale channel models and concepts of \emph{time cluster spatial lobes} (TCSL) to model multiple multipath time clusters that are seen to arrive in particular directions~\cite{Rap15a,Rap13a,Rap15b,Samimi15a,Samimi15b,Samimi15c,Nie13a}.

This paper presents a brief overview of the indoor channel properties for bands up to 100 GHz based on extensive measurements and results across a multitude of bands. In addition we present a preliminary set of channel parameters suitable for indoor 5G simulations that are capable of capturing the main properties and trends.

\section{Requirements For New Channel Model}

The requirements of the new channel model that will support 5G operation across frequency bands up to 100 GHz are outlined below:
\begin{enumerate}
\item The new channel model should preferably be based on the existing 3GPP 3D channel model~\cite{3GPP36873} but with extensions to cater for additional 5G modeling requirements and scenarios, for example:
	\begin{enumerate}
		\item Antenna arrays, especially at higher-frequency millimeter-wave bands, will very likely be 2D and dual-polarized both at the access point (AP) and at the user equipment (UE) and will hence need properly-modeled azimuth and elevation angles of departure and arrival of multipath components.
		\item Individual antenna elements will have antenna radiation patterns in azimuth and elevation and may require separate modeling for directional performance gains.  Furthermore, polarization properties of the multipath components need to be accurately accounted for in the model.  
	\end{enumerate}
\item The new channel model must accommodate a wide frequency range up to 100 GHz.  The joint propagation characteristics over different frequency bands will need to be evaluated for multi-band operation, e.g., low-band and high-band carrier aggregation configurations. 
\item The new channel model must support large channel bandwidths (up to 2 GHz), where:
	\begin{enumerate}
		\item The individual channel bandwidths may be in the range of 100 MHz to 2 GHz and may support carrier aggregation.
		\item The operating channels may be spread across an assigned range of several GHz.
	\end{enumerate}
\item The new channel model must support a range of large antenna arrays, in particular:
	\begin{enumerate}
		\item Some large antenna arrays will have very high directivity with angular resolution of the channel down to around 1.0 degree.
		\item 5G will consist of different array types, e.g., linear, planar, cylindrical and spherical arrays, with arbitrary polarization.
		\item The array manifold vector can change significantly when the bandwidth is large relative to the carrier frequency. As such, the wideband array manifold assumption is not valid and new modeling techniques may be required. It may be preferable, for example, to model departure/arrival angles with delays across the array and follow a spherical wave assumption instead of the usual plane wave assumption.
	\end{enumerate}
\item The new channel model must accommodate mobility, in particular (for outdoor models, although mentioned here for consistency):
	\begin{enumerate}
		\item The channel model structure should be suitable for mobility up to 350 km/hr.
		\item The channel model structure should be suitable for small-scale mobility and rotation of both ends of the link in order to support scenarios such as device to device (D2D) or vehicle to vehicle (V2V).
	\end{enumerate}
\item The new channel model must ensure spatial/temporal/frequency consistency, in particular:
	\begin{enumerate}
		\item The model should provide spatial/temporal/frequency consistencies which may be characterized, for example, via spatial consistence, inter-site correlation, and correlation among frequency bands. 
		\item The model should also ensure that the channel states, such as line-of-sight (LOS)/non-LOS (NLOS) for outdoor/indoor locations, the second order statistics of the channel, and the channel realizations change smoothly as a function of time, antenna position, and/or frequency in all propagation scenarios. 
		\item The spatial/temporal/frequency consistencies should be supported for simulations where the channel consistency impacts the results (e.g. massive MIMO, mobility and beam tracking, etc.).  Such support could possibly be optional for simpler studies.
	\end{enumerate}
\item The new channel model must be of practical computational complexity, in particular:
	\begin{enumerate}
		\item The model should be suitable for implementation in single-link simulation tools and in multi-cell, multi-link radio network simulation tools. Computational complexity and memory requirements should not be excessive. The 3GPP 3D channel model~\cite{3GPP36873} is seen, for instance, as a sufficiently accurate model for its purposes, with an acceptable level of complexity. Accuracy may be provided by including additional modeling details with reasonable complexity to support the greater channel bandwidths, and spatial and temporal resolutions and spatial/temporal/frequency consistency, required for millimeter-wave modeling.
		\item The introduction of a new modeling methodology (e.g. Map based model) may significantly complicate the channel generation mechanism and thus substantially increase the implementation complexity of the system-level simulator. Furthermore, if one applies a completely different modeling methodology for frequencies above 6 GHz, it would be difficult to have meaningful comparative system evaluations for bands up to 100 GHz.
	\end{enumerate}
\end{enumerate}

\section{Indoor Deployment Scenarios - Indoor (InH): Open and Closed Office, Shopping Mall}
\begin{figure}[b!]
	\centering
	\includegraphics[width = 3.7in]{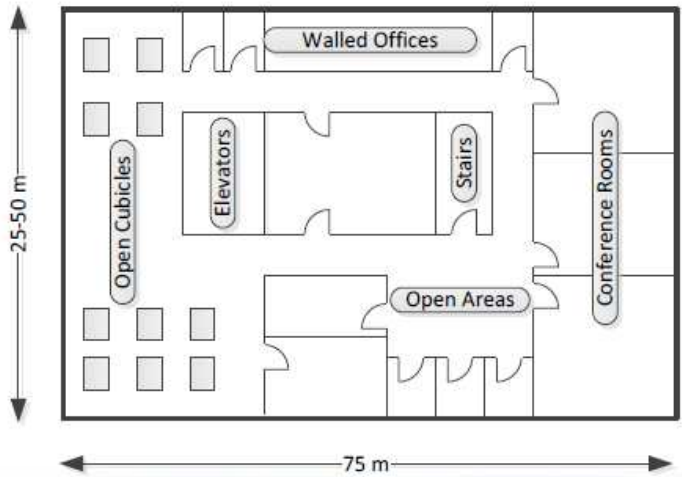}
	\caption{Typical Indoor Office.}
	\label{fig:InHsc}
\end{figure}

\begin{figure}[b!]
	\centering
	\includegraphics[width = 3.7in]{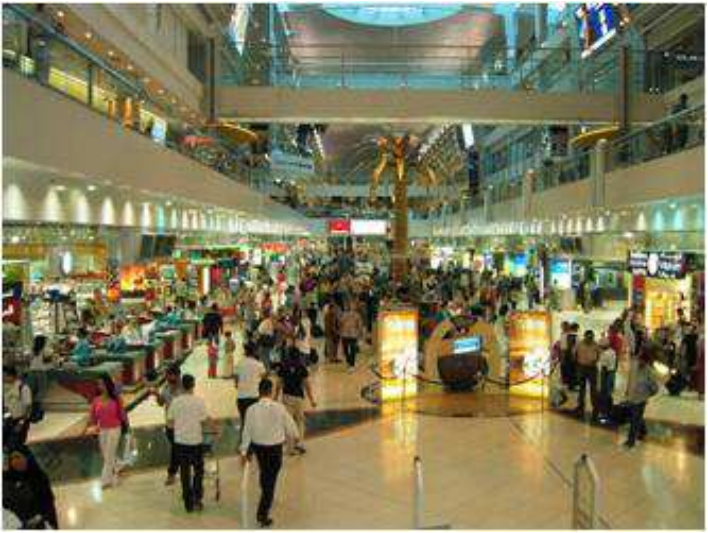}
	\caption{Indoor Shopping Malls.}
	\label{fig:SMsc}
\end{figure}

The indoor scenario includes open and closed offices, corridors within offices and shopping malls as examples. The typical office environment has open cubicle areas, walled offices, open areas, corridors, etc., where the partition walls are composed of a variety of materials like sheetrock, poured concrete, glass, cinder block, etc. For the office environment, APs are generally mounted at a height of 2-3 m either on the ceilings or walls, with UEs at heights between 1.2 and 1.5 m. Shopping malls are generally 2-5 stories high and often include an open area (``atrium"). In the shopping-mall environment, APs are generally mounted at a height of approximately 3 m on the walls or ceilings of the corridors and shops, with UEs at heights between 1.2 and 1.5 m. The density of APs may range from one per floor to one per room, depending on the frequency band and output power. A typical indoor office and shopping mall scenario are shown in Figures~\ref{fig:InHsc} and~\ref{fig:SMsc}, respectively. 

\section{Characteristics of the InH Channel from 6 GHz to 100 GHz}
Measurements over a wide range of frequencies have been performed by the co-authors of this paper. In the following sections we outline the main observations per scenario with some comparisons to the existing 3GPP models for below 6 GHz (e.g.~\cite{3GPP36873}).  

In LOS conditions, multiple reflections from walls, floor, and ceiling give rise to waveguiding. Measurements in both office and shopping mall scenarios show that path loss exponents, based on a 1 m free space reference distance, are typically below 2 in LOS conditions, leading to more favorable path loss than predicted by Friis' free space path loss formula. The strength of the waveguiding effect is variable and the path loss exponent appears to increase very slightly with increasing frequency, possibly due to the relation between the wavelength and surface roughness. 

Measurements of the small scale channel properties such as angular spread and delay spread have shown remarkable similarities between channels over a very wide frequency range. It appears as if the main multipath components are present at all frequencies though with some smaller variations in amplitudes.

Recent work shows that polarization discrimination ranges between 15 and 25 dB for indoor millimeter wave channels~\cite{Karttunen15a}, with greater polarization discrimination at 73 GHz than at 28 GHz~\cite{Mac15a}.

\section{Penetration Inside Buildings}
\begin{figure}
	\centering
	\includegraphics[width = 3.7in]{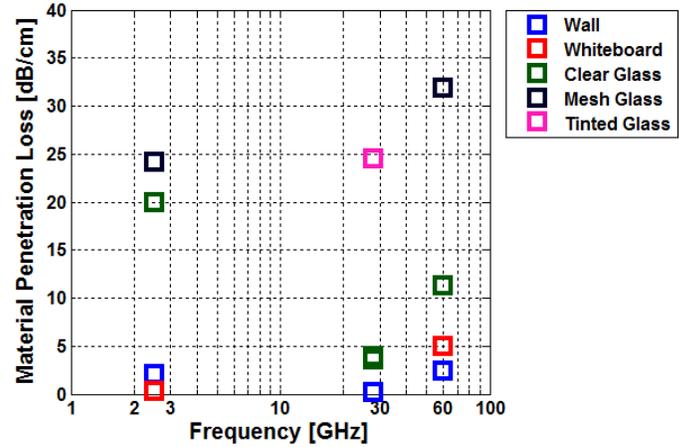}
	\caption{2.5 GHz, 28 GHz, and 60 GHz normalized material penetration losses from indoor measurements with common types of glass and walls were lumped together into common datasets~\cite{Rap15b,Anderson02a,Zhao13a}.}
	\label{fig:NYUpenetration}
\end{figure}

Measurements have been reported for penetration loss for various materials at 2.5, 28, and 60 GHz for indoor scenarios~\cite{Rap15a,Rap13a,Anderson02a,Zhao13a}, although all materials were not measured for the same frequencies. For easy comparisons, walls and drywalls were lumped together into a common dataset and different types of clear class were lumped together into a common dataset with normalized penetration loss shown in Figure~\ref{fig:NYUpenetration}. It was observed that clear glass has widely varying attenuation (20 dB/cm at 2.5 GHz, 3.5 dB/cm at 28 GHz, and 11.3 dB/cm at 60 GHz). For mesh glass, penetration was observed to increase as a function of frequency (24.1 dB/cm at 2.5 GHz and 31.9 dB/cm at 60 GHz), and a similar trend was observed with whiteboard penetration increasing as frequency increased. At 28 GHz, indoor tinted glass resulted in a penetration loss of 24.5 dB/cm. Walls showed very little attenuation per cm of distance at 28 GHz (less than 1 dB/cm)~\cite{5GSIG}. Furthermore, a simple parabolic model as a function of frequency for low-loss and high-loss building penetration is given in~\cite{5GSIG_VTC16}.

\section{Path loss, Shadow Fading, LOS, and Blockage Modeling}
\subsection{LOS Probability}
The definition of LOS used in this paper is discussed in this sub-section together with other LOS models.  The LOS state is determined by a map-based approach, i.e., by considering the transmitter (AP) and receiver (UE) positions and whether any buildings or walls block the direct path between the AP and the UE. The impact of objects not represented in the map such as chairs, desks, office furniture, etc. is modelled separately using shadowing/blocking terms. An attractive feature of this LOS definition is that it is frequency independent, as only walls are considered in the definition.  

\begin{figure}
	\centering
	\includegraphics[width = 3.7in]{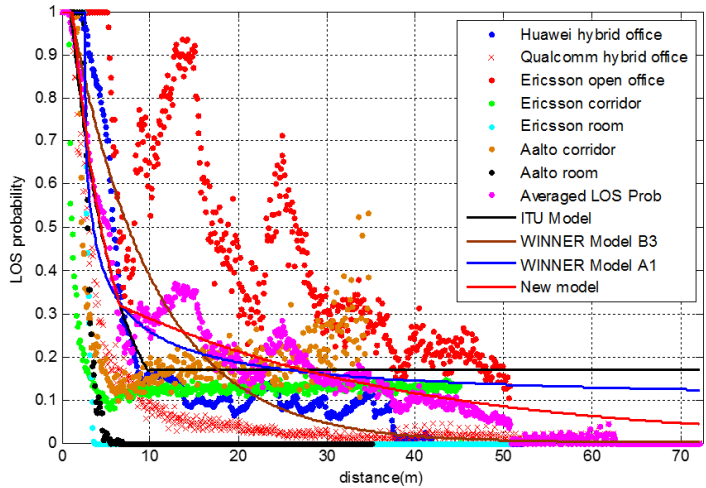}
	\caption{Indoor office LOS probability for the three models considered.}
	\label{fig:InHpLOS}
\end{figure}

Since the 3GPP 3D model~\cite{3GPP36873} does not include an indoor scenario for LOS-probability, and the indoor hotspot scenario in e.g. the IMT advanced model~\cite{ITU-M.2135-1} differs from the office scenario considered in this paper, an investigation on the LOS probability for indoor office has been conducted based on ray-tracing simulations. Different styles of indoor office environments were investigated, including open-plan office with cubical area, closed-plan office with corridor and meeting room, and also a hybrid-plan office with both open and closed areas. It has been verified that the following model fits the propagation in indoor office environment the best, of the three models evaluated:
\begin{equation}\label{eq1}
P_{LOS} = \begin{cases}
1, & d\leq1.2\text{ m}\\
\exp(-(d-1.2)/4.7), & 1.2<d<6.5\text{ m}\\
\exp(-(d-6.5)/32.6)\cdot 0.32, & d\geq6.5\text{ m}
\end{cases}
\end{equation}
The verification results are shown in Table~\ref{tbl:InHpLOS} and Figure~\ref{fig:InHpLOS}. The LOS probability model used in ITU IMT-Advanced evaluation~\cite{ITU-M.2135-1} and WINNER II~\cite{WINNERII} are also presented here for comparison. For the ITU and WINNER II model, parameterization results based on new data vary a lot from the original model. The results show that the new model has a good fit to the data in an average sense and can be used for 5G InH scenario evaluations. However, note the high variability between different deployments and degrees of openness in the office area.
\begin{table*}
\caption{Comparison of the LOS probability models for the InH environment}\label{tbl:InHpLOS}
\centering
\renewcommand{\arraystretch}{1.6}
\begin{center}
	\scalebox{1}{
		\fontsize{8}{8}\selectfont
		\begin{tabular}{|c||c|c|c|}
			\hline
			Models	& Original	& Updated/New	& MSE \\ \hline \hline
			ITU Model	& \parbox{6.5cm}{\begin{equation*}\label{eq2} P_{LOS} = \begin{cases} 1, & d\leq18\text{ m}\\ \exp(-(d-18)/27), & 18\text{ m}<d<37\text{ m}\\ 0.5, & d\geq37\text{ m} \end{cases} \end{equation*}} & \parbox{6.5cm}{\begin{equation*}\label{eq3} P_{LOS} = \begin{cases} 1, & d\leq1.1\text{ m}\\ \exp(-(d-1)/4.9), & 1.1\text{ m}<d<9.8\text{ m}\\ 0.17, & d\geq9.8\text{ m} \end{cases} \end{equation*}} & 0.0499 \\ \hline
			\scriptsize WINNER II model (B3)	& \parbox{6.5cm}{\begin{equation*}\label{eq4} P_{LOS} = \begin{cases} 1, & d\leq10\text{ m}\\ \exp(-(d-10)/45), & d>10\text{ m}\end{cases}\end{equation*}} & \parbox{6.5cm}{\begin{equation*}\label{eq5} P_{LOS} = \begin{cases} 1, & d\leq1\text{ m}\\ \exp(-(d-1)/9.4), & d>1\text{ m}\end{cases}\end{equation*}} & 0.0572  \\ \hline
			\scriptsize WINNER II model (A1)	& \parbox{6.7cm}{\begin{equation*}\tiny\label{eq6} P_{LOS} = \begin{cases} 1, & d\leq2.5\text{ m}\\ 1-0.9(1-(1.24-0.61\log_{10}(d))^3)^{1/3}, & d>2.5\text{ m}\end{cases}\end{equation*}} & \parbox{6.7cm}{\begin{equation*}\tiny\label{eq7} P_{LOS} = \begin{cases} 1, & d\leq2.6\text{ m}\\ 1-0.9(1-(1.16-0.4\log_{10}(d))^3)^{1/3}, & d>2.6\text{ m}\end{cases}\end{equation*}} & 0.0473  \\ \hline
			\scriptsize New Model	& N/A & \parbox{7.5cm}{\begin{equation*}\label{eq8}
			P_{LOS} = \begin{cases} 1, & d\leq1.2\text{ m}\\ \exp(-(d-1.2)/4.7), & 1.2\text{ m}<d<6.5\text{ m}\\ \exp(-(d-6.5)/32.6)\cdot 0.32, & d\geq6.5\text{ m} \end{cases} \end{equation*}} & 0.0449 \\ \hline
		\end{tabular}}
	\end{center}
\end{table*}

\subsection{Path Loss Models}
To adequately assess the performance of 5G systems, multi-frequency path loss (PL) models, LOS probability, and blockage models will need to be developed across the wide range of frequency bands and for operating scenarios. Three PL models are considered in this paper; namely the close-in (CI) free space reference distance PL model~\cite{Rap15b,Anderson02a,Sun15a} the close-in free space reference distance model with frequency-dependent path loss exponent (CIF)~\cite{Mac15a}, and the Alpha-Beta-Gamma (ABG) PL model~\cite{Mac15a,Piersanti12a,Hata80a,Mac13a}. These models are described in the following text and are then applied to various scenarios.

Table~\ref{tbl:InHPL} shows the parameters of the CI, CIF, and ABG path loss models for different environments for omnidirectional antennas. It may be noted that the models presented here are multi-frequency models, and the parameters are invariant to carrier frequency and can be applied across the 0.5-100 GHz band.
\begin{equation}\label{eq9}
\PL^{CI}(f,d)[\dB]=\FSPL(f,1\;\text{m})+10n\log_{10}\left(\frac{d}{1\;\text{m}}\right)+X_{\sigma}^{CI}
\end{equation}
where $f$ is the frequency in Hz, $n$ is the PLE, $d$ is the distance in meters, $X_{\sigma}^{CI}$ is the shadow fading (SF) with $\sigma$ in dB, and the free space path loss (FSPL) at 1 m, with frequency $f$ is given as:
\begin{equation}\label{eq10}
\FSPL(f,1\;\text{m})=20\log_{10}\left(\frac{4\pi f}{c}\right),
\end{equation}
where $c$ is the speed of light.

The ABG PL model~\cite{Sun16a,Thomas16a,Piersanti12a,Mac15a} is given as:
\begin{equation}\label{eq11}
\begin{split}
\PL^{\ABG}(f,d)[\dB]=10\alpha\log_{10}(d)+\beta\\
+10\gamma\log_{10}(f)+X^{\ABG}_{\sigma}
\end{split}
\end{equation}
where $\alpha$ captures how the PL increase as the transmit-receive distance (in meters) increases, $\beta$ is a floating offset value in dB, $\gamma$ attempts to capture the PL variation over the frequency $f$ in GHz, and $X^{\ABG}_{\sigma}$  is the SF term with standard deviation in dB.

The CIF PL model is an extension of the CI model~\cite{Mac15a}, and uses a frequency-dependent path loss exponent given by:\begin{equation}\label{eq9}
\begin{split}
\PL^{\CIF}(f,d)[\dB]=\FSPL(f,\text{1 m})+\\10n\Bigg(1+b\left(\frac{f-f_0}{f_0}\right)\Bigg)\log_{10}\left(\frac{d}{\text{1 m}}\right)+X^{\CIF}_{\sigma}
\end{split}
\end{equation}
where $n$ denotes the path loss exponent (PLE), and $b$ is an optimization parameter that captures the slope, or linear frequency dependency of the path loss exponent that balances at the centroid of the frequencies being modeled (e.g., path loss increases as $f$ increases when $b$ is positive). The term $f_0$ is a fixed reference frequency, the centroid of all frequencies represented by the path loss model~\cite{Mac15a}, found as the weighed sum of measurements from different frequencies, using the following equation:
\begin{equation}\label{eq10}
f_0 = \frac{\sum_{k=1}^Kf_k N_K}{\sum_{k=1}^K N_K}
\end{equation}
where $K$ is the number of unique frequencies, and $N_k$ is the  number of path loss data points corresponding to the $k^{th}$  frequency $f_k$. The input parameter $f_0$ represents the weighted frequencies of all measurement (or Ray-tracing) data applied to the model. The CIF model reverts to the CI model when $b$ = 0 for multiple frequencies, or when a single frequency $f$ = $f_0$ is modelled. For InH, a dual-slope path loss model might provide a good fit for different distance zones of the propagation environment. Frequency dependency is also observed in some of the indoor measurement campaigns conducted by co-authors. For NLOS, both a dual-slope ABG and dual-slope CIF model can be considered for 5G performance evaluation (they each require 5 modeling parameters to be optimized), and a single-slope CIF model (that uses only 2 optimization parameters) may be considered as a special case for InH-Office~\cite{Mac15a}. The dual-slope may be best suited for InH-shopping mall or large indoor distances (greater than 50 m). The dual slope InH large scale path loss models are given in~\eqref{eq11} (ABG) and~\eqref{eq12} (CIF).
\begin{figure*}[t!]
	\begin{equation}\label{eq11}
	\resizebox{0.9\hsize}{!}{$%
		\PL_{Dual}^{ABG}(f,d)[\dB] = \begin{cases} \alpha_1\cdot10\log_{10}(d)+\beta_1+\gamma\cdot10\log_{10}(f), & 1\text{ m}<d\leq d_{BP}\\ \alpha_1\cdot10\log_{10}(d_{BP})+\beta_1+\gamma\cdot10\log_{10}(f)+\alpha_2\cdot 10\log_{10}\left(\frac{d}{d_{BP}}\right), & d>d_{BP}
	\end{cases}$}
	\end{equation}
\end{figure*}
\begin{figure*}
	\begin{equation}\label{eq12}
	\resizebox{0.94\hsize}{!}{$%
		\PL_{Dual}^{CIF}(f,d)[\dB] = \begin{cases} \FSPL(f,\text{1 m})+10n_1\left(1+b_1\left(\frac{f-f_0}{f_0}\right)\right)\log_{10}\left(\frac{d}{\text{1 m}}\right), & 1\text{ m}<d\leq d_{BP}\\
		\FSPL(f,\text{1 m})+10n_1\left(1+b_1\left(\frac{f-f_0}{f_0}\right)\right)\log_{10}\left(\frac{d_{BP}}{\text{1 m}}\right)+10n_2\left(1+b_2\left(\frac{f-f_0}{f_0}\right)\right)\log_{10}\left(\frac{d}{d_{BP}}\right), & d>d_{BP}
		\end{cases}$}
	\end{equation}
	\hrulefill
	% The spacer can be tweaked to stop underfull vboxes.
	\vspace*{4pt}
\end{figure*}

In the CI PL model, only a single parameter, the path loss exponent (PLE), needs to be determined through optimization to minimize the SF standard deviation over the measured PL data set~\cite{Rap15b,Sun15a,Sun16a}. In the CI PL model there is an anchor point that ties path loss to the FSPL at 1 m, which captures frequency-dependency of the path loss, and establishes a uniform standard to which all measurements and model parameters may be referred. In the CIF model there are 2 optimization parameters ($n$ and $b$), and since it is an extension of the CI model, it also uses a 1 m free-space close-in reference distance path loss anchor. In the ABG PL model there are three parameters which need to be optimized to minimize the standard deviation (SF) over the data set~\cite{Mac15a,Sun16a}. Closed form expressions for optimization of the model parameters for the CI, CIF, and ABG path loss models are given in~\cite{Mac15a}, where it was shown that indoor channels experience an increase in the PLE value as the frequency increases, whereas the PLE is not very frequency dependent in outdoor UMa or UMi scenarios~\cite{Mac15a,Rap15b,Sun15a,Sun16a,Thomas16a}. The CI, CIF, and ABG models, as well as cross-polarization forms and closed-form expressions for optimization are given for indoor channels in~\cite{Mac15a}.

Another important issue related to path loss is shadow fading. For InH, the distance dependency and frequency dependency were investigated for both indoor office and shopping mall. For the LOS propagation condition, the frequency and distance dependency is weak. But for the NLOS propagation condition, frequency and distance dependency is more apparent as indicated in Table 7 of~\cite{5GSIG}.

\begin{table}
	\caption{InH and Shopping Mall Path Loss Models for LOS and NLOS.}\label{tbl:InHPL}
	\centering
	\renewcommand{\arraystretch}{1.6}
	\begin{center}
	\scalebox{0.82}{
		\fontsize{8}{8}\selectfont
		\begin{tabular}{|p{3cm}|p{3cm}|p{3cm}|}
			\hline
			Scenario & CI/CIF Model Parameters & ABG Model Parameters \\ \hline \hline
			InH-Indoor-Office-LOS & $n$=1.73, $\sigma_{\SF}$=3.02 dB & N/A \\ \hline
			InH-Indoor-Office-NLOS single slope (FFS)  & $n$=3.19, $b$=0.06, $f_0$=24.2 GHz, $\sigma_{\SF}$=8.29 dB & $\alpha$=3.83, $\beta$=17.30, $\gamma$=2.49, $\sigma_{\SF}$=8.03 dB \\ \hline
			InH-Indoor-Office-NLOS dual slope & $n_1$=2.51, $b_1$=0.12, $f_0$=24.1 GHz, $n_2$=4.25, $b_2$=0.04, $d_{BP}$=7.8 m, $\sigma_{\SF}$=7.65 dB & $\alpha_1$=1.7, $\beta_1$=33.0, $\gamma$=2.49, $d_{BP}$=6.90 m, $\alpha_2$=4.17, $\sigma_{\SF}$=7.78 dB \\ \hline
			InH-Shopping Malls-LOS & $n$=1.73, $\sigma_{\SF}$=2.01 dB & N/A \\ \hline
			InH-Shopping Malls-NLOS single slope (FFS)  & $n$=2.59, $b$=0.01, $f_0$=39.5 GHz, $\sigma_{\SF}$=7.40 dB & $\alpha$=3.21, $\beta$=18.09, $\gamma$=2.24, $\sigma_{\SF}$6.97 dB \\ \hline
			InH-Shopping Malls-NLOS dual slope & $n_1$=2.43, $b_1$=0.01, $f_0$=39.5 GHz, $n_2$=8.36, $b_2$=0.39, $d_{BP}$= 110 m, $\sigma_{\SF}$=6.26 dB & $\alpha_1$=2.9, $\beta_1$=22.17, $\gamma$=2.24, $d_{BP}$=147.0 m, $\alpha_2$=11.47, $\sigma_{\SF}$=6.36 dB \\ \hline
		\end{tabular}}
	\end{center}
\end{table}

\section{Fast Fading Modeling}
For InH scenarios, an investigation of fast fading modelling has been conducted based on both measurement and ray-tracing. Both indoor office and shopping mall environments have been investigated at frequencies including 2.9 GHz, 3.5 GHz, 6 GHz, 14 GHz, 15 GHz, 20 GHz, 28 GHz, 29 GHz, 60 GHz, and 73 GHz. Some preliminary analysis on large-scale channel characteristics have been summarized in~\cite{5GSIG}. Although it is still too early to apply these results to the full frequency range up to 100 GHz, these preliminary investigations have provided insight into the difference induced by the largely extended frequency range. The preliminary analysis in~\cite{5GSIG} illustrates the frequency dependency of large-scale channel characteristics across the measured frequency range.
\section{Conclusion}
The basis for this paper is the open literature in combination with recent and ongoing propagation channel measurements performed by a majority of the co-authors of this paper, some of which are as of yet unpublished. The InH propagation models are somewhat different from the outdoor UMi and UMa models in that the indoor channels are more frequency-dependent than outdoor channels, leading to the ABG and CIF frequency-dependent NLOS path loss models. In LOS conditions, waveguiding effects were observed in all frequencies measured, leading to path loss exponents less than the theoretical value of $n$ = 2 in LOS. The preceding tables give an overview of these recent measurement activities in different frequency bands and scenarios, in addition to further information provided in~\cite{5GSIG}.
\section*{Acknowledgment}
The authors would like to thank Jianhua Zhang\textsuperscript{b} and Yi Zheng\textsuperscript{c} who are also contributing authors of this manuscript and 5G white paper~\cite{5GSIG}.

\bibliographystyle{IEEEtran}
\bibliography{5GCMSIG_ICC_V1_2_arxiv}
\end{document}